# Vesicle-like structure of lipid-based nanoparticles as drug delivery system revealed by molecular dynamics simulations

Maryam Khalkhali[a], Sarah Mohammadinejad[b]*, Farhad Khoeini[a]*, Kobra Rostamizadeh[c]

[a]Department of Physics, University of Zanjan, 45195-313, Zanjan, IRAN

[b]Department of Biological Sciences, Institute for Advanced Studies in Basic Sciences (IASBS), Zanjan 45137-66731, IRAN

[c]Zanjan Pharmaceutical Nanotechnology Research Center, Zanjan University of Medical Sciences, 45139-56184, Zanjan, IRAN

## Abstract
Lipid-based drug delivery systems are considered as promising vehicles for hydrophobic drug compounds. Lipid distribution within the droplet can affect drug loading capacity in these carriers. However, it is extremely challenging to determine the nanostructure within these carriers through the implementation of the direct experimental methods due to the ultrafine size. Therefore, coarse grained molecular dynamics (MD) simulation was utilized to model different kinds of lipid-based nanoparticles of the diameter about 12 nm including solid lipid nanoparticles (SLN), nanoemulsion (NE), and nanostructured lipid carriers (NLC), and the organization of the lipids within the carriers was explored. The aforementioned nanoparticles consisted of stearic acid, oleic acid as lipids, and sodium dodecyl sulfate (SDS) as a surfactant in water medium. Furthermore, the impact of solid to liquid mass ratio on the lipid distribution within the lipid matrix was investigated regarding the NLC simulations. In the equilibrium state, we observed the vesicle-like structure for all the investigated systems in which the hydrophilic moieties of the lipids and surfactant organized a semi-bilayer fold into the droplet and the hydrophobic tails accumulated among them. It is worth mentioning although SDS as a harsh surfactant, which is a special case, was expected to be present in the surface of the droplet, it penetrated into the lipids. Additionally, our results showed remarkable entrapped water beads inside the droplet in the form of one or more cavities along the internal layer of the head groups which was surrounded by lipid head groups. It was also reported that in the building structure of the nanoemulsion and SLN, in the central parts of the droplets, lipids were denser than the case of NLCs. Moreover, no crystallization occurred within the lipid-based carriers. Finally, the results indicated that, in the case of NLC simulations, the lipid distribution within the lipid matrix was insensitive to the mass fraction of solid to liquid lipids.

## 1. Introduction
Nowadays, lipid-based nanocarriers have received extensive attention as desirable delivery systems from scholars and researchers [1-4]. Among these lipid-based formulations, solid lipid nanoparticles (SLN), in which lipid matrix is composed of one solid lipid or a blend of solid lipids, have been developed as a capable vehicle for the encapsulation and controlled release of pharmaceuticals and lipophilic compounds. Although it can be very efficient in protecting the chemically unstable compounds against degradation [5, 6], there are also reported disadvantages associated with SLNs, including highly ordered recrystallization of the structure after cooling, which results in lower encapsulation, poor controlled release as well as physical instability [7, 8]. Accordingly, nanostructured lipid carriers (NLC) with the coexistence of solid and liquid lipid matrix were developed to deal with these issues (with preferably solid to liquid ratio in the range of 70:30 up to 99.9:0.1) [9]. It was supposed that the existence of liquid lipid inside the nanoparticle would inhibit highly ordered recrystallization after the preparation process leading to an increased drug loading capacity and improved release profile [10]. Therefore, nanostructured lipid carriers (NLC) have obtained an increased interest because of various advantages such as the increased drug loading capacity [11, 12], controlled drug release [13], high drug stability [14], biocompatibility, and non-toxicity [15, 16].

Several investigations concerning the physicochemical properties of SLN and NLC have been carried out so far. As an instance, Chen et al. investigated the possibility of using NLCs to improve the oral absorption of lovastatin [17]. For the purpose of their study, NLCs consisting of Precirol and squalene were constructed, and the effective properties of the carriers were compared with both SLNs composed of pure Precirol and lipid emulsions (LEs) composed of squalene. They found that more than 70% of the drug was trapped in NLCs and LEs, which was remarkably higher than SLNs. They also reported that various Precirol/squalene ratios could modify the lipid matrix and subsequently affect the drug release.

Based on the obtained results, it is probable that the internal structure of NLC plays a prominent role in specifying these carriers' performance. Therefore, it is essential to get some insight into these internal structures and determine the drug distribution in these types of delivery systems [18]. However, to the best of our knowledge, there is no precise and accepted information on the internal structure of such systems at different conditions since routinely available measuring techniques are not capable of determining these complicated internal structures.

Despite the lack of sufficient experimental equipment and methodologies, some studies suggested different spatial arrangements for the solid/liquid domains in NLCs. As an example, Jores et al. [19] obtained a great deal of detailed information on physicochemical properties of SLN and NLC experimentally. They applied various analytical methods to evaluate the structures and lipid dispersions inside SLN and NLC based on glyceryl behenate and medium chain triglycerides. They did not find NLCs spherical droplets similar to nanoemulsions but detected them as lipid platelets with sticky oil spots on the surface. They also proved that in the case of crystalline lipid-based carriers, the increased water-lipid interfaces cause low drug linkage in comparison with nanoemulsions. Furthermore, Jores et al. [20], utilized spectrofluorometry and Raman spectroscopy to demonstrate that NLCs are partly solid platelets with oil, available between the solid platelet and the surfactant layer. Tikekar et al. [21] measured the distribution of an entrapped lipid soluble dye in SLNs and NLCs using fluorescence imaging. Results demonstrated that due to solidification of lipids in SLNs, the entrapped dye was expulsed into the aqueous phase. However, the existence of oil in NLCs prevented the exclusion of encapsulated dye and recaptured the dye within the liquid domains restricted within the solid phase.

Additionally, electron paramagnetic resonance (EPR) spectroscopy has been utilized in order to provide sufficient detailed information on the organization of the lipid core of the nanoparticles as well as drug distribution within the lipid-based carriers. In this regard, Barbosa et al. [22] reported the presence of lamellar arrangement inside the lipid nanoparticles applying EPR measurement, which was confirmed by small angle X-ray diffraction. In another study, Haag et al. [23] used EPR spectroscopy to determine TEMPO (2, 2, 6, 6-tetramethyl-1-piperidinyloxy) distribution within the NLC and evaluated the dynamics of the skin penetration. W-band EPR spectroscopy revealed 35% TEMPO within the lipid compartments of the nanostructured lipid carriers.

Recently, Boreham et al. [18] used a novel procedure based on fluorescence microscopy to determine drug distribution precisely and visualize the internal structure of a lipophilic fluorescent drug loaded NLCs. The results of the study revealed a spherical drug loaded NLCs composed of liquid nanocompartment. The calculation of diffusion constant proved that the drug molecule is distributed in the liquid component, and the oily domain fills about 10% to 50% of the NLCs' volume. As another example, Selvamuthukumar et al. [24] classified nanostructured lipid carriers into three types. In type one, solid lipids and liquid lipids are mixed, forming extremely disordered defective lipid matrix constructions. Type two has high oil concentrations in which a miscibility gap between solid lipid and liquid lipid occurs during the chilling step that lead to phase separation and form oily nanocompartments. In this type, drug can be placed in the solid part, but at excessed solubility, it can be distributed in the oily parts of the structure. Finally, in type three, the lipid matrix consists of solid and liquid lipids with a noncrystalline structure. This amorphous structure avoids drug expulsion which can be attributed to the liquid phase.

According to the several experimental documents achieved by different excipients and production process, various observations about the internal structure of lipid-based carriers would be supposedly expected. The hypothesis predicts different distributions such as a homogeneous distribution of the liquid and solid lipids within the structure, a heterogeneous structure where domains accommodate similar to lipid bilayers, a core−shell structure that the liquid lipids build the shell on the surface of the solid phase or vice versa. Liquid lipid nanocompartments are also placed within the solid phase or on the surface of the solid phase [18, 25].

In addition to the previously mentioned experimental evaluations, molecular dynamics simulations (MD) are potential to serve as a powerful technique to provide detailed information about the internal structure of nanostructured drug carriers and drug localization within them. There are just a few molecular dynamics studies focusing on systems like lipid droplets (LDs) and the distribution of lipids in such systems.

Chaban et al. [26] employed the coarse-grained molecular dynamics simulation method within the MARTINI framework to characterize the molecular structure of lipid droplets composed of triolein and cholesteryl oleate with various fractions of cholesteryl oleate to triolein. These LDs were covered by a phospholipid monolayer formed by a mixture of 1-palmitoyl-2-oleoylphosphatidylcholine (POPC) and 1-palmitoyl-2-oleoyl-sn-glycero-3-phosphoethanolamine (POPE) phospholipid molecules. Their simulation results demonstrated that cholesteryl oleate molecules are frequently located within the hydrophobic core of lipid droplets, and their penetration in the monolayer is small. In other words, tirolein and cholesteryl oleate retain a single phase, which forms a hydrophobic core of the lipid droplets.

In another study, Chaban and Khandelia [27] used coarse-grained molecular dynamics (MD) simulations to investigate lipid droplets formation in systems including triolein, cholesterol, POPC, and POPE in water medium. They found that cholesterol often accumulates in the central part of the lipid droplets instead of the interface. Additionally, they reported that the presence of PE lipids at the interface does not have an impressive effect on the internal structure of lipid droplets and the distribution of excipients.

Benson et al. [28] utilized the molecular dynamics simulations to examine the effect of excipients composition on droplet nanostructure and drug localization within the self-emulsifying drug delivery systems (SEDDS). They designed a standard SEDDS system including capric triglycerides and cyclosporin A as the drug in aqueous solution and investigated the influence of variation of drug content, fatty acid chain length, the addition of surfactant and surfactant concentration on droplet nanostructure as well as drug localization within it. The simulation results showed that the alteration of fatty acid chain lengths led to forming different recognizable droplet patterns such as lamellar-like, random, and vesicle-like. Besides, they reported that the addition of poly (ethylene glycol) (PEG-6) as a surfactant does not have any remarkable result on the droplet internal nanostructure.

As the review of related literature presents, few available MD studies directly focus on the distribution of lipids within the lipid-based nanocarriers. Owing to this fact, in the current research, coarse grained molecular dynamics (MD) simulation was used to study the lipid-based nanocarriers comprising stearic acid (solid lipid), oleic acid (liquid lipid), and sodium dodecyl sulfate (SDS) as surfactant in water medium. The effect of solid to liquid ratio on the spatial distribution of the lipid matrix was also investigated. Depending on the mass fraction of oleic acid to stearic acid, eleven different compositions constructed. After equilibration, the internal structures of the equilibrated lipid-based carrier obtained and diverse characterization analyses such as radius of gyration, radial distribution functions, and radial density distribution were utilized to characterize and determine the internal structure of the lipid-based nanocarriers.

## 2. Methodology

### 2.1. Simulation Systems

The MD simulation was carried out to examine the effect of solid to liquid mass ratio on the equilibrium internal arrangement of lipids inside the lipid-based nanocarriers. To achieve this purpose, lipid-based nanocarriers were constructed as droplets of the diameter about 12 nm and with 11 different solid to liquid ratios for various simulations. According to the previous experiment [29], stearic and oleic acid were chosen as the solid lipid and liquid lipid respectively. The initial structures of the droplets were constructed in a way that stearic acid and oleic acid molecules were randomly placed inside it. The droplet was put within a cubic box of length 21 nm and then surrounded by randomly distributed water molecules, sodium dodecyl sulfate (SDS) as the surfactant and neutralizing $Na^+$ ions. In all of the systems, the number of water molecules, SDS and ions were the same. The initial configuration of the systems was created by using the Packmol software package [30].

The complete list of all eleven simulated systems and the details of MD simulation are obviously presented in table 1. System 1 (ST100) is the simulation of SLN, consisting of only stearic acid (solid lipid), systems 2−10 (STX-OLY) are NLCs which include oleic acid (liquid lipid) and stearic acid (solid lipid) at different proportions. System 11 (OL100) is a nanoemulsion (NE) which contains only oleic acid (liquid lipid).

Table 1. Summary of simulated systems

| Simulation name | No. Stearic acid | No. Oleic acid | No. SDS | No. $Na^+$ ions | No. water | No. CG beads | Simulation time (μs) |
|---|---|---|---|---|---|---|---|
| 1: ST100 | 1770 | 0 | 400 | 400 | 265000 | 77100 | 12 |
| 2: ST90-OL10 | 1593 | 177 | 400 | 400 | 265000 | 77100 | 12 |
| 3: ST85-OL15 | 1504 | 266 | 400 | 400 | 265000 | 77100 | 12 |
| 4: ST80-OL20 | 1416 | 354 | 400 | 400 | 265000 | 77100 | 12 |
| 5: ST75-OL25 | 1327 | 443 | 400 | 400 | 265000 | 77100 | 12 |
| 6: ST70-OL30 | 1239 | 531 | 400 | 400 | 265000 | 77100 | 12 |
| 7: ST60-OL40 | 1062 | 708 | 400 | 400 | 265000 | 77100 | 12 |
| 8: ST50-OL50 | 885 | 885 | 400 | 400 | 265000 | 77100 | 12 |
| 9: ST40-OL60 | 708 | 1062 | 400 | 400 | 265000 | 77100 | 12 |
| 10: ST30-OL70 | 531 | 1239 | 400 | 400 | 265000 | 77100 | 12 |
| 11: OL100 | 0 | 1770 | 400 | 400 | 265000 | 77100 | 12 |

The topology files for stearic acid and SDS molecules were downloaded from http://cgmartini.nl/index.php/ force-field-parameters. The model of oleic acid was built out of the default MARTINI atom types in accordance with DOPC tail (http://cgmartini.nl/ndex.php/force-field-parameters/lipids2/351-lipid.html?dir=PC&lipid=DOPC) since it has the similar tail with DOPC. The schematic view of oleic acid molecule mapping has been fully depicted in Fig. 1. As illustrated, this model is made up of a single polar head group particle connected to a chain of four lipid tail particles. All bonded particles have been kept together with a harmonic potential with a force constant of $K_{bond}$=1250 kJ mol$^{-1}$ nm$^{-2}$ and an equilibrium distance of $R_{bond}$ = 0.47 nm. Moreover, to take chain bending stiffness into account, a harmonic potential of the cosine type with a force constant of $k_{angle}$ = 25 kJ mol$^{-1}$ and an equilibrium bond angle of $\Theta_0$ = 180° was used. It is notable that due to the existence of cis unsaturated bond, for the angle between $C_1$, $C_3$, and $C_1$ beads the force constant is set to 45 kJ mol$^{-1}$ and an equilibrium bond angle is set to 120° [31].

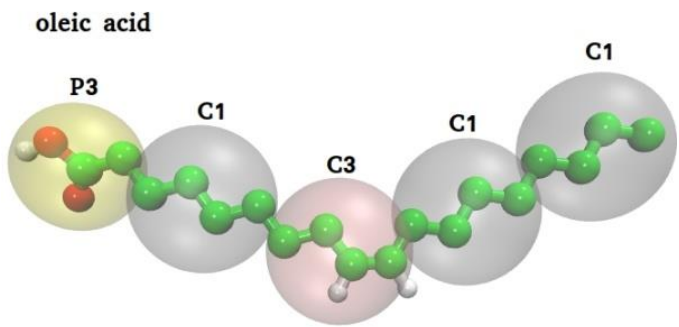

Fig. 1. Schematic view of the coarse grained model used for an oleic acid molecule.

## 2.2. Simulation Details

Simulations were carried out using GROMACS 5.0.4 package [32, 33] within the MARTINI 2.1 framework [31, 34] with Intel(R) Xeon(R) CPU E5-2690 v2 @ 3.00GHz with 40 cores. They were performed in the NPT ensemble with the pressure set to 1 bar and temperature annealed from 353 to 300 K in 3 steps (at the first step heating up to 353 K in 4 μs, then upholding the temperature at 353 K in 3 μs and finally, cooling down to 300 K in 5 μs). The temperature and pressure were connected to baths according to the method presented by Berendsen et al. [35]. With the thermostat and barostat relaxation times of 1.0 and 3.0 ps, respectively [27], periodic boundary conditions were applied in three dimensions. Electrostatic and Lennard-Jones interactions were shifted respectively from 0.0 to 1.2 nm and from 0.9 to 1.2 nm in order to smoothly decay to zero [36, 37]. The leapfrog integrator [38] with a time step of 30 fs was used in production MD. The translational motion of the droplet center of mass was eliminated every 30 fs, and it was kept in the middle of the simulation box. As it has been illustrated in table 1, all systems were simulated for the entire12 μs. The first 7 μs were discarded as the pre-equilibration step.

**3. Results and Discussion**

A lipid-based drug carrier was modeled by coarse grained molecular dynamics simulations to find the spatial distribution of the lipid matrix and the impact of the mass fraction of solid to liquid lipids on the internal nanostructure. Starting with an initial randomly mixed droplet of diameter 12 nm and running the simulation dynamics for12 μs for all different systems (table 1), we carried out several structural analyses on the equilibrated droplets.

Fig. 2 represents the equilibrated configurations of lipid-based droplets with different mass fractions of solid to liquid lipids. The snapshots were captured from the last time frames of solid lipid nanoparticles (SLN), nanoemulsion (NE), and different ratios of lipids in the nanostructured lipid carriers (NLC). As expected, due to the fact that amphiphilic molecules are known to perform a bias of their hydrophilic parts toward water molecules, in the final configurations, the amphiphilic moieties (stearic acid, oleic acid, and SDS) formed a vesicle-like structure in which their hydrophilic heads were exposed to water and the hydrophobic tails were located between them. All the snapshots show complete miscibility of the head groups of the two liquid and solid lipids as well as SDS. Furthermore, on the surface of the droplet, the head group beads are in direct contact with water beads.

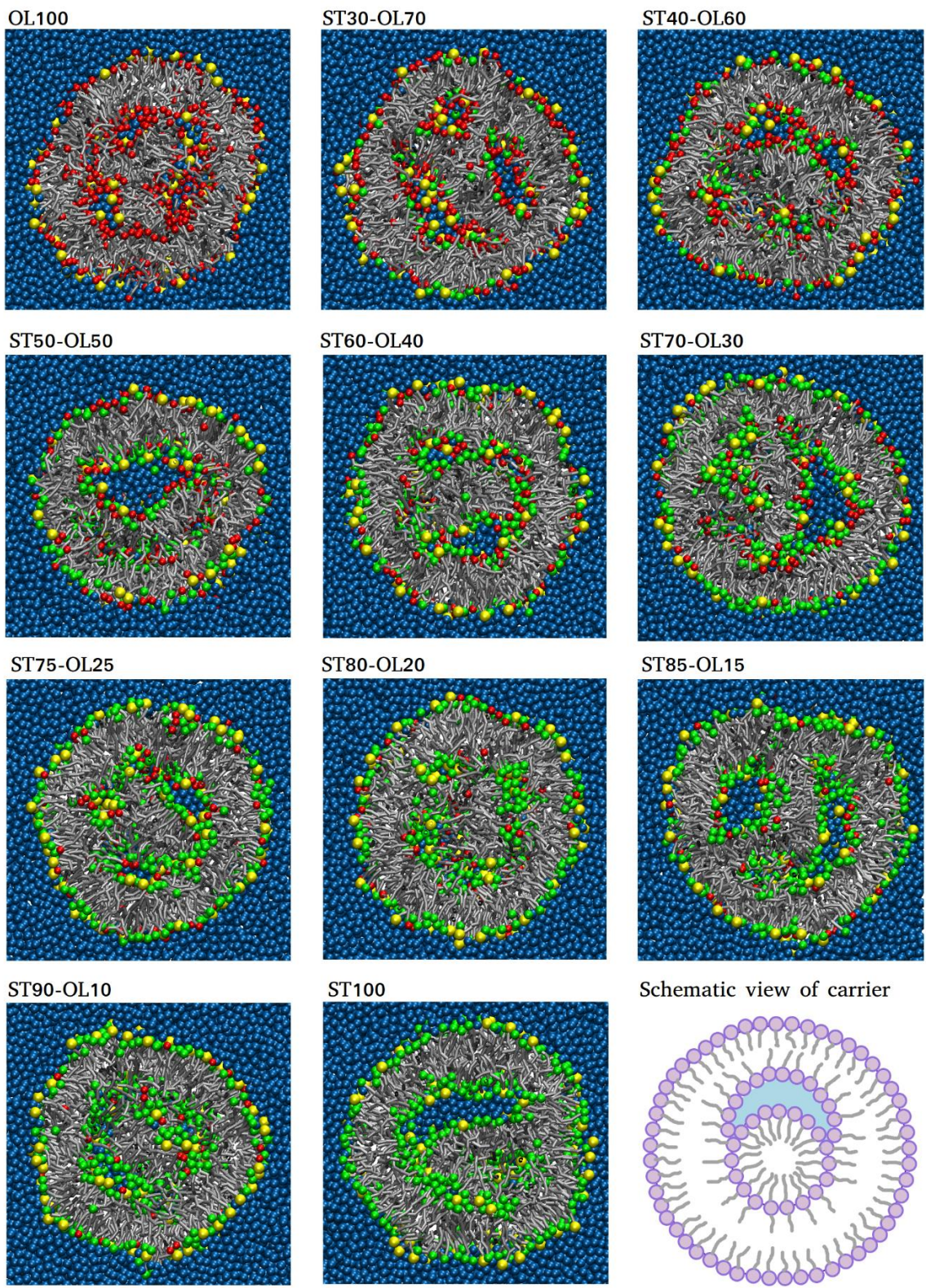

Fig. 2. Final distribution of lipid moieties inside droplets at different solid to liquid lipid fractions. The snapshots are captured from the cross-section of droplets at last time frames of simulation. Stearic acid head group are rendered in green, oleic acid head group in red, SDS head group in yellow, water beads in blue and all the hydrophobic tails in gray.

### 3.1. Radius of gyration

The radius of gyration of each lipid-based carrier was defined as the root mean-square distance of the beads to its center-of-mass to evaluate both the stability of the system and the effect of lipid fraction on droplet equilibrium size. Fig. 3(a) displays the radius of gyration versus time for different solid to liquid lipid fractions, that is for OL100 (black line), ST30-OL70 (violet line), ST50-OL50 (blue line), ST70-OL30 (red line), ST85-OL15 (pink line) and ST100 (green line) systems during the time after equilibration. For all systems shown in this figure, it can be observed that the radius of gyration remains stable after equilibrium. Furthermore, the nanoemulsion (OL100) and SLN (ST100) systems in which the entire droplet is respectively composed of a uniform liquid and solid lipids exhibit somewhat the smallest radii of gyration. However, the St50-OL50 system that is the combination of both liquid and solid lipids with equal proportion attains the largest radius.

Moreover, the radius of the droplet can be calculated by finding the distance of the farthest atoms from the droplet center of mass. Fig. 3(b) shows the droplet radius versus time for different solid to liquid lipid fractions, that is for OL100 (black line), ST30-OL70 (violet line), ST50-OL50 (blue line), ST70-OL30 (red line), St85-Ol15 (pink line) and ST100 (green line) systems during the time after equilibration. As it is obvious, the radius of the droplet remains stable at an approximate value of 8 nm after the equilibrium. Furthermore, in contrast to the radius of gyration, all the simulations with different lipid ratios show the same radius for the droplets.

We observed the same radius for the droplets in all lipid fractions while the radii of gyration for OL100 and ST100 were smaller than that of the other fractions. This observation can be interpreted in a way that the lipid distribution is denser in the central part of the OL100 and St100 simulations in comparison with other systems. This point may be the advantage of NLCs to NEs and SLNs because the denser distribution of lipids in the center of the droplet can reduce its drug loading capacity due to the crystallization effects. As mentioned earlier, previous studies demonstrated that lipid crystallization inside nanostructure lipid carriers reduces their drug loading capacity. Besides, the existence of liquid lipid inside the nanoparticle would prevent the highly ordered crystallization after preparation, and therefore increase the drug loading capacity of the NLCs [10].

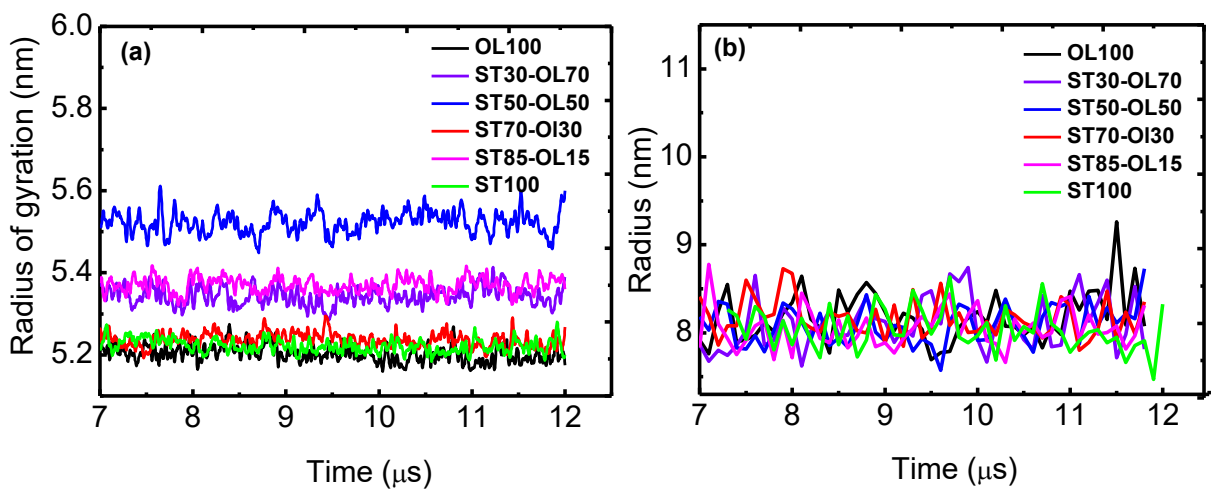

Fig. 3. a) Radius of gyration and b) radius of droplets versus time for different solid to liquid lipid fractions, OL100 (black line), ST30-OL70 (violet line), ST50-OL50 (blue line), ST70-OL30 (red line), ST85-OL15 (pink line) and ST100 (green line) systems after system equilibration.

## 3.2. Radial distribution function

The radial distribution function (RDF) depicts how the density of one molecule type changes as a function of distance from another reference molecule type in the system. It can be exploited to explore the details of the structural orders of the molecules in the droplet.

In this study, the radial distribution function for different pairs of molecule types were calculated in the simulated systems. Since it is not possible to present the calculated RDFs for all the simulated systems, only the calculated RDFs for ST70-OL30 system was presented as this ratio is one of the preferred solid to liquid ratios for NLCs according to the available studies on NLCs [9]. Fig. 4 illustrates RDFs calculated for different pairs of molecules in the ST70-OL30 system. Oleic head-oleic head RDF plot (solid black line) in Fig. 4(a) depicts some peaks. The first one located at 0.52 nm is the most prominent one, proposing that oleic head groups are located as close as 0.52 nm to each other within the droplets. The second peak is located at 0.92 nm which is less remarkable than the first one.

Similarly, stearic head-stearic head RDF plot (red dashed line) demonstrates the same results for stearic acid distribution. In the case of SDS head groups (green dash-dotted line), two peaks are observed at the same locations as above, i.e., at 0.52 and 0.92. The first peak at 0.52 nm is lower than the peak at 0.92 nm. Observing the second peak as the dominant peak implies that SDS head groups are not as compact as oleic and stearic head groups. Moreover, the first peak for SDS is smaller than those of oleic acid and stearic acid due to the low concentration of SDS in comparison to lipid molecules.

Fig. 4(b) shows the calculated RDFs for oleic head groups-stearic head groups (solid black line), oleic head groups-SDS head groups (red dashed line), and stearic head groups-SDS head groups (green dash-dotted line). Three RDFs have almost the same trends as their first and prominent peak at 0.52 nm with the same height. These similar trends for the distribution function of cross-pairs of stearic and oleic and SDS molecules indicate the miscibility of the head groups of the two liquid and solid lipids as well as SDS in agreement with the snapshots of Fig. 2. This may also suggest the formation of the polar islands inside the droplets by gathering the head groups of the molecule head groups close to each other.

It is worth stating that RDFs in Fig. 4 (a-b) resembles the typical RDF plots representing the orders of a liquid phase. Based on this similarity, it can be concluded that crystallization did not occur in the droplets [26, 27].

Fig. 4(c) illustrates three RDFs between water, lipids, and SDS head groups. It is found that lipids and SDS head groups can approach closer to the water beads because the first peak is located at 0.52. Moreover, considering SDS head groups, the height of the first peak is higher than the lipids implying that in comparison, SDS has the greater tendency to the water molecules. This observation confirms the detergent role of SDS molecules in the interface of the droplet with water environment for stabilization purposes.

Fig. 4(d) depicts three RDFs between water, lipids, and SDS tail groups. The results indicate that in contrast to the head groups, tails tend to keep away from water molecules. SDS tails have the highest first peak and tend to be closer to the water molecule compared with the tails of other molecules. Again this higher tendency for SDS tails toward water molecules can be attributed to the surfactant feature of SDS molecules.

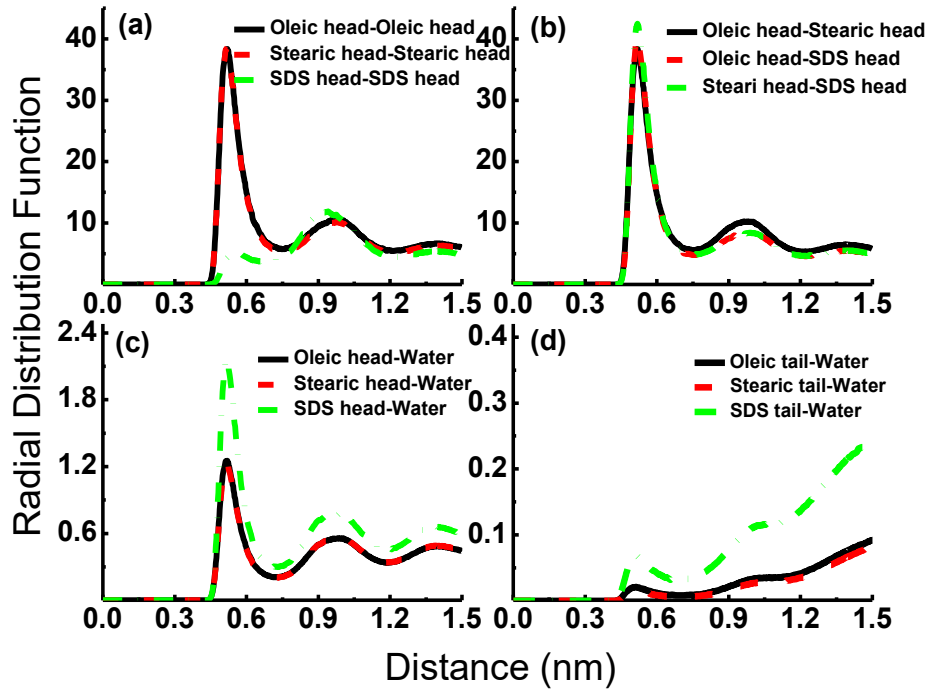

Fig. 4. Radial distribution functions between selected moieties at ST70-OL30. (a) Oleic head groups-oleic head groups (solid black line), stearic head groups-stearic head groups (red dashed line), and SDS head groups-SDS head groups (green dash-dotted line). (b) Oleic head groups-stearic head groups (solid black line), oleic head groups-SDS head groups (red dashed line), stearic head groups-SDS head groups (green dash-dotted line), (c) Oleic head groups-water (solid black line), stearic head groups-water (red dashed line), SDS head groups-water (green dash-dotted line). (d) Oleic tails-water (solid black line), stearic tails-water (red dashed line), and SDS tails-water (green dash-dotted line).

### 3.3. Density distribution function

In order to fully understand the packing of the molecules in the droplets, the radial density of head groups of all moieties versus their distance from the droplet center of mass was calculated. Fig. 5 (a-d) shows the radial density profile for oleic, stearic, and SDS head groups for various solid to liquid fractions. It is observed that in all the systems with different solid to liquid fractions, head groups of lipids have the same density profiles. Additionally, two dominant peaks were found in all cases; one is nearly located at the radius of 3.6 nm and the latter at 6.5 nm indicating that most of the head groups are placed in these two radii. The larger radius (6.5 nm) almost coincides with the radius of the droplet representing the accumulation of polar head groups in the outside of the droplets due to their hydrophobic nature and water tendency. This two-peak density profile is reminiscent of the profile density of the lipid polar groups in the vesicular structures and can be interpreted as the vesicle-like structure adopted by the droplet.

It is also worth mentioning that altering the solid to liquid ratio would change the height of the peaks without any switch in their location. It was expected that the systems with different solid to lipid ratios have different density profiles. However, the results show the same density profiles for different lipid ratios which may be attributed to the fact that the same numerical parameters were used for coarse graining of stearic acid and oleic acid head groups in the performed coarse grained force field.

Furthermore, the observed peaks of density profiles for SDS molecules are nearly located in the same position as the peaks for lipid head groups, which illustrates certain solubility of lipids in the surfactant. Similarly, the density profile of SDS molecules is not sensitive to the solid to the liquid fraction and is the same in all of the cases. Comparing the height of the two peaks for SDS and lipid head groups reveals that the density of lipid head groups in the inner region of the droplet is more than the outer region while both peaks have almost the same density in SDS. This can serve justification for the special role of SDS molecules. SDS is a harsh detergent which is often used for membrane protein isolation from their native membranes by intercalating into the lipid bilayers. Although it is expected that SDS exists at the surface of the droplet, it penetrates into the lipids, which is in line with the results obtained by Deo et al. [39].

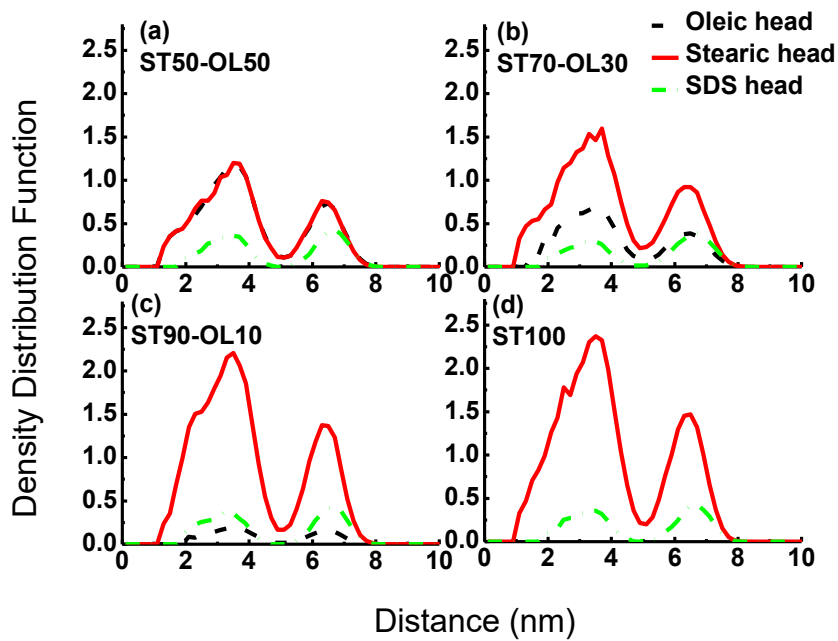

Fig. 5. Radial density profile of head group beads of lipids and SDS molecules as a function of the distance from the droplet center of mass in different solid to lipid ratio simulations.

Fig. 6 represents the radial density profile for the tail beads of stearic acid, oleic acid, and SDS molecules in addition to the water beads versus their distance from the droplet center of mass. It should be stated that only the last beads of the tails for all the molecules were taken into consideration. It can be seen that the tails of both lipids have the same patterns for their tail density profiles. Two peaks can be observed in all cases, one is almost in the central part of the droplet and the second one is nearly in 5 nm from the droplet center. Considering the most probable location for the presence of head groups from Fig. 5, it can be concluded that most of the tails are located in between the regions occupied by the hydrophilic moieties which is the prominent feature of the vesicular structure. The location of the first peak which is very close to the center of the droplet indicates that unlike the perfect vesicular structure in which the center is filled with water, hydrophobic region is in the center of the droplet.

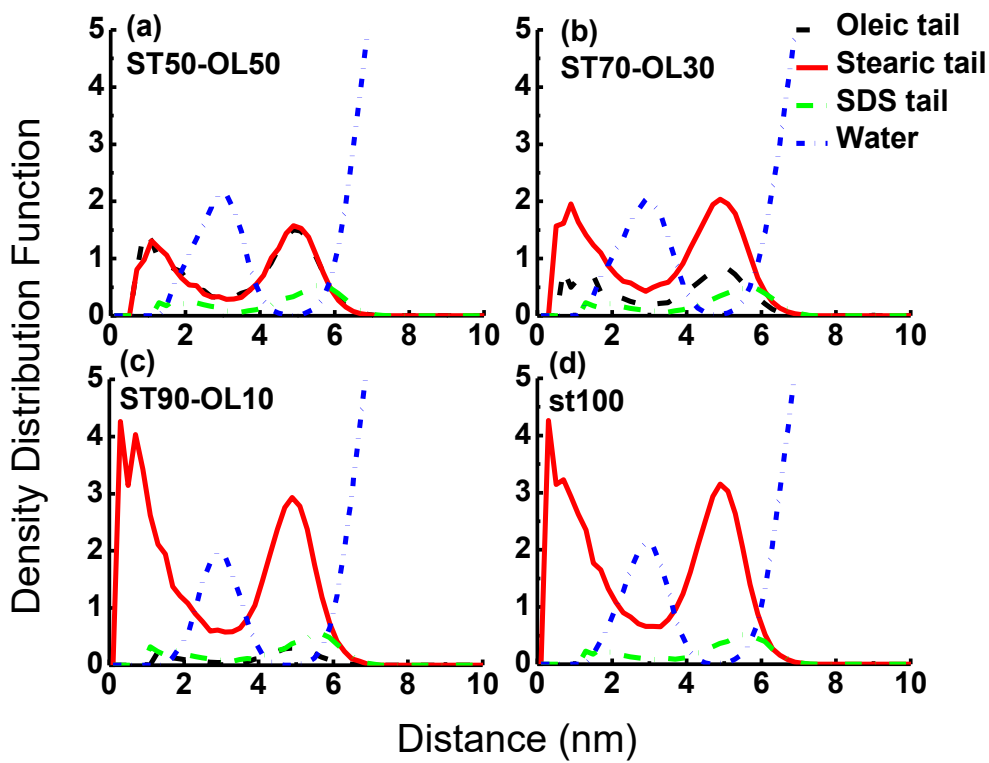

Fig. 6. Radial distribution of the number density of tail beads of lipids and SDS and water beads in different simulated systems as a function of the distance from the droplet center of mass.

Looking at water density profiles (Fig. 6) also implies that there is a remarkable entrapped water region inside all the simulated systems, and the peak is located approximately at 2.9 nm from the droplet center. The mentioned peak is close to the first density peak for the head groups at 3.6 nm. Considering the locations of the density peaks and the snapshots in Fig. 2, it is concluded that the entrapped water region is formed of one or more cavities filled with water inside the internal layer of the head groups and surrounded by the head groups. The previous interpretation can be shown schematically by drawing a simplified view of the droplet as shown in Fig. 2.

In lipid-based systems such as solid lipid nanocarriers (SLN), nanoemulsion (NE), and nanostructured lipid carriers (NLC), a homogenous lipidic composition has been widely accepted. However, having a vesicle-like structure for lipid-based nanoparticles is of great importance in terms of providing the possibility to carry hydrophilic as well as hydrophobic drugs.

Moreover, in the tail density profiles, it is observed that in the case of ST90-OL10 and ST100 systems the first peak is both higher and closer to the droplet center of mass, indicating that lipids are denser in the central parts of these two simulated systems. Possessing higher density in the center of SLNs was also predicted with the results of the radius of gyration analyses.

## 4. Conclusions

In this study, coarse grained molecular dynamics (MD) simulation was utilized within the MARTINI framework to investigate the internal structure of the various types of lipid-based nanocarriers such as solid lipid nanocarriers (SLN), nanoemulsion (NE), and nanostructured lipid carrier (NLC) composed of stearic acid, oleic acid, and SDS as solid lipid, liquid lipid, and surfactant in water medium. Therefore, taking NLC simulations into consideration, the effect of solid to liquid mass ratio on the lipid distribution within the lipid matrix was evaluated. The calculation of radius and gyration radius for all simulations demonstrated that the distribution of lipids in the central parts of the nanoemulsion and SLN are denser than the case of NLCs.

RDF and radial density calculations demonstrated that all of the investigated systems in the present study formed the vesicle-like structure with the hydrophilic head groups of the lipids and surfactant organized a semi-bilayer fold into the droplet, and the hydrophobic tails accumulated among them. Water density profile indicated that there is a remarkable entrapped water region inside the droplet in the form of one or more cavities along the internal layer of the head groups surrounded by lipid head groups. A simplified representation of the droplet internal structure (shown in Fig. 2) revealed a vesicle-like structure which is altered and deviated in its central part. The water region is shifted to one side of the droplet and lets a part of the hydrophobic tail accommodate in the center of the vesicle.

Furthermore, the results provided that no crystallization occurs within the lipid-based carriers. Moreover, it was surprisingly figured out that the lipid distribution within the lipid matrix is insensitive to the fraction of solid to liquid mass, but due to the formation vesicle-like structure and the existence of water beads in the inner parts of the droplet, this lipid-based nanocarrier can be proposed as a promising vehicle for the delivery of both hydrophilic and hydrophobic drugs.

Previous studies prove that various factors affect the features of the NLCs, including lipid ratios, type of lipids and surfactant, preparation method, homogenization stirring speed and duration, etc. It is accepted that these factors alter the load and release capacity of NLCs by changing the internal distribution of lipids inside them. In the present study, the influence of lipid ratio on the internal structure of lipid-based nanoparticles was investigated. Regarding the reliable results of the MARTINI force field, apparently altering the fraction of stearic acid to oleic acid did not significantly change the droplet's internal structure.

Due to the fact that stearic acid and oleic acid are very similar in their structure, in the coarse graining process, very similar numerical parameters are assigned to both, especially in their head groups. This may be the reason for not illustrating the considerable difference in the internal structures by changing the lipid ratios. Therefore, some further investigations on the droplets with other types of solid and liquid lipids can be suggested by keeping in mind choosing solid and liquid lipids with more difference in their atomic structure so that their coarse grained structures differ more remarkably.

Furthermore, as far as the authors are concerned, most experimentally produced NLCs are reported to have sizes around a hundred nanometers. In this research, due to the limited power of the computer facilities, NLCs were modeled as a sphere of diameter 12 nm which is smaller than the reported size. In future, there is an opportunity to test if simulating the droplets with the bigger size closer to real NLCs would affect their internal structure.

Finally, it is assumed that the internal structure of the NLCs would not change remarkably in the presence of the drugs, i. e. the internal structure can be generalized to the case of the droplets including the drugs. However, to complete these investigations, the researchers propose to include the drug molecules inside the droplets and investigate the extent to which their assumptions come true. Besides, it is feasible to look for important details such as drugs localization in the lipid-base carriers and accordingly predict their load and release features.

b*E-mail: khoeini@znu.ac.ir

a*E-mail: sarah@iasbs.ac.ir